\documentclass[onecolumn,preprint,preprintnumbers,amsmath,amssymb]{revtex4}

\usepackage{graphicx}
\usepackage{dcolumn}
\usepackage{bm}

\begin{document}
\title{Magnetically induced signatures in a thin film superconductor}

\author{Lars Egil Helseth}
\address{Max Planck Institute of Colloids and Interfaces, D-14424 Potsdam, Germany}%

\begin{abstract}
We study the interaction between a one dimensional magnetic nanostructure and a thin film superconductor. 
It is shown that different magnetic distributions produce characteristic magnetic field signatures. 
Moreover, the magnetic structure can induce a weak link in the superconducting film, or be positioned directly 
above a predefined nonsuperconducting weak link. We estimate the magnetic flux associated with such a
structure, and discuss a general expression for the energy calculated within the London model.

\end{abstract}

\pacs{Valid PACS appear here}
\maketitle

Weak links in superconductors have generated a lot of interest over the past decades, both in the study of 
conventional and high $T_{c}$ superconductors. More recently, research has also been focused towards 
junctions in thin films, due to their potential in future technology (see e.g. Ref. \cite{Kogan} 
and references therein). For such junctions it has been found that 
the field is a superposition of fields from Pearl vortices along the junction with a certain line density\cite{Kogan}. 
An interesting special case is the generation of weak links with spatially localized magnetic fields.
Creation of a weak link in a bulk (or thick film) superconductor using a magnetic domain wall was first proposed 
by Sonin\cite{Sonin}. In that paper expressions for the magnetic fields where found, and 
it was argued that domain walls can induce movable weak links. In the current paper we try to extend the idea of 
Ref. \cite{Sonin} to thin film superconductors, and estimate the magnetic field and flux distribution 
associated with an one dimensional magnetic nanostructure. The weak link could be a domain wall, generated and 
controlled by a stress pattern or external field, or it could be a stationary prefabricated nanomagnetic stripe.
A particular advantage of using magnetic domain walls as weak links is that they can be 
moved at high speeds, and may therefore have potential applications in future fluxtronics devices. On the other hand, in 
a prefabricated magnetic stripe (e.g. permalloy) the polarity of the magnetic vector could easily be switched by 
an external field\cite{Yi}. This could also be of interest for creation and annihilation of vortices. 
Since the vortex pinning energy often can be regarded as proportional to the thickness of the superconductor, it 
is reasonable to assume that pinning by nonmagnetic sources are negligible in a thin film. Therefore, it should in 
principle be possible to produce film systems in which the magnetic texture determines the junction properties.

Consider a thin superconducting film located at 
$z=0$ with thickness much smaller than the penetration depth of the
superconductor. The surface is covered by a one dimensional magnetic structure centered at $x=0$, with 
thickness much smaller than that of the superconducting film. It is assumed 
that the magnetic structure is sufficiently long that end effects are not a problem, and that it consists of 
surface charges separated from the superconductor by a very 
thin (negligible) oxide layer to avoid the exchange of electrons or spin (see e.g. Ref. \cite{Erdin}). 
We will here analyze the case where the magnetization is perpendicular to the plane of the superconducting film, since 
we expect this geometry to give a stronger coupling to the superconductor. In order to gain some insight into the behavior of the weak link, let us first compare the magnetic fields from 
the two following models for the magnetization distribution:
\begin{equation}
\mbox{\boldmath $M$}^{G} = M_{0} \exp (-\alpha x^{2}) \delta(z)\mbox{\boldmath $\hat{e}$}_{z} \,\,\, , 
\end{equation}
and
\begin{displaymath}
\mbox{\boldmath $M$}^{S} = \left\{ \begin{array}{ll}
M_{0}\delta(z)\mbox{\boldmath $\hat{e}$}_{z} & \textrm{if  $-W \leq x \leq
W$}\\
0 & \textrm{if $|x|  \geq W$}\\  
\end{array} \right. ,
\end{displaymath}
where $M_{z0}$ and $\alpha$ are constants. Here we assume that $\alpha =2W^{-2}$. 

Using the method Ref. \cite{Helseth,note1}, we find that the gaussian magnetization distribution results in the following 
field components (per unit length):
\begin{equation}
H^{G}_{z}(x,z) =\frac{\lambda _{e} M_{0}}{\sqrt{\pi \alpha} }\int_{0}^{\infty}k_{x} ^{2} \frac{\exp\left( -k_{x}^{2}/4\alpha \right) cos(k_{x}x)}{1+2\lambda _{e}k_{x}}
\exp(-k_{x}|z|)dk_{x} \,\,\, ,
\label{Hz}
\end{equation}
\begin{equation}
H^{G}_{x}(x,z) =\frac{\lambda _{e} M_{0}}{\sqrt{\pi \alpha} }\int_{0}^{\infty}k_{x} ^{2} \frac{ \exp\left( -k_{x}^{2}/4\alpha  \right) sin(k_{x}x)}{1+2\lambda _{e}k_{x}}
\exp(-k_{x}|z|)dk_{x}
\,\,\, .
\label{Hx}
\end{equation}
On the other hand, the step magnetization distribution gives 
\begin{equation}
H^{S}_{z}(x,z) =\frac{2\lambda _{e} M_{0}}{\pi}\int_{0}^{\infty}k_{x}  \frac{sin(k_{x}W) cos(k_{x}x)}{1+2\lambda _{e}k_{x}}
\exp(-k_{x}|z|)dk_{x} \,\,\, ,
\label{Hz}
\end{equation}
\begin{equation}
H^{S}_{x}(x,z) =\frac{2\lambda _{e} M_{0}}{\pi }\int_{0}^{\infty}k_{x}\frac{ sin(k_{x}W) sin(k_{x}x)}{1+2\lambda _{e}k_{x}}
\exp(-k_{x}|z|)dk_{x}
\,\,\, .
\label{Hx}
\end{equation}

\begin{figure}
\includegraphics[width=12cm]{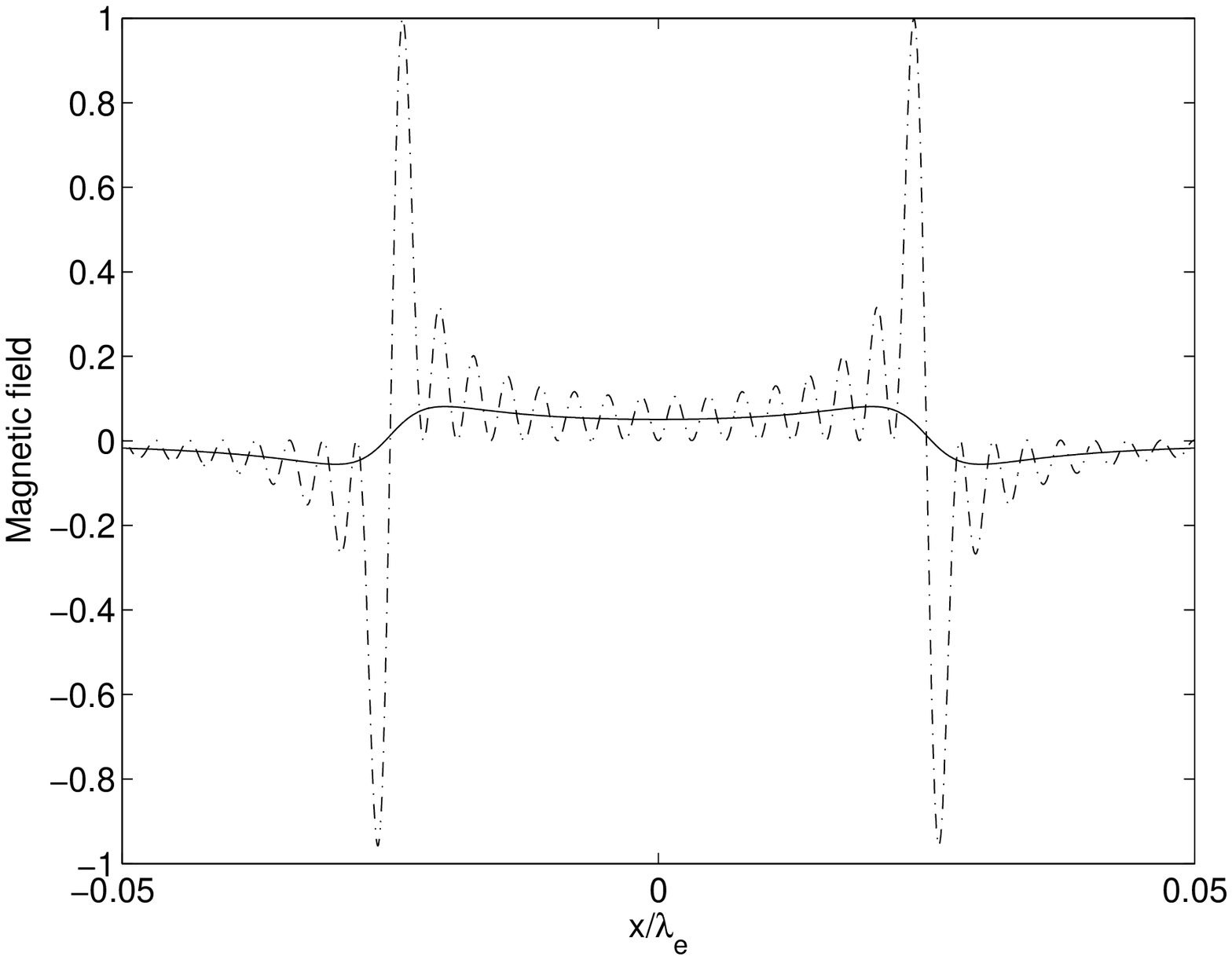}
\caption{\label{f1}  The z component of the magnetic field generated by a step-like magnetization distribution 
when $z=0$ (dash-dotted line) and $z=\lambda _{e}/200$ (solid line). Here $W=\lambda /40$, and the curves
have been normalized with respect to the maximum peak of the dash-dotted line.}
\vspace{2cm}
\end{figure}

\begin{figure}
\includegraphics[width=12cm]{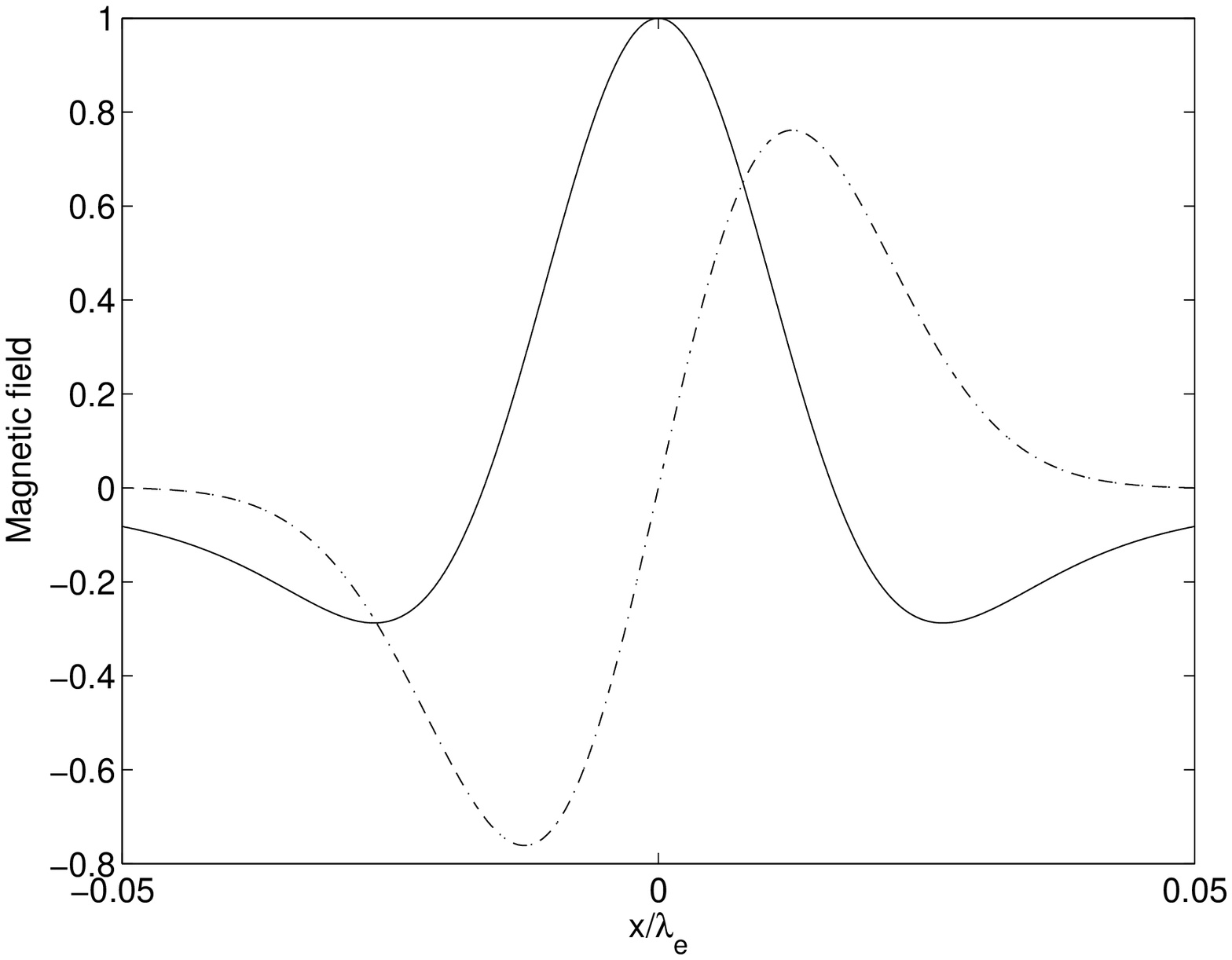}
\caption{\label{f2}  The x (dash-dotted line) and z (solid line) component of the magnetic field generated by a 
gaussian magnetization distribution. Here z=0, $W=\lambda /40$ and $\alpha =2W^{-2}$. The curves are normalized with
respect to the maximum peak of the z component.}
\vspace{2cm}
\end{figure}

Figure \ref{f1} shows $H^{S}_{z}$ when $z=\lambda _{e}/200$ (solid line) and $z=0$ (dash-dotted line). Note that  
when $z=0$ the magnetic field oscillates due to the steep magnetization gradient. This could therefore be interpreted as 
Gibbs oscillations, well known in Fourier analysis. At a certain height above the surface these oscillations are smoothed out due to the exponential 
decay factor. 
Figure \ref{f2} shows $H^{G}_{z}$ (solid line) and $H^{G}_{x}$ (dash-dotted line) when $W=\lambda _{e}/40$ and $z=0$. 
Note that the peak of the z component is located at the origin. 
This is in contrast to Fig. \ref{f1}, where the maximum field is located near 
the edges. Moreover, the two negative peaks of the field are much less pronounced for a gaussian distribution. 

It is clear that if the magnetic field exceed the critical field of the superconductor, then a weak link is 
generated at which vortices may exists\cite{Sonin}. The current across the link is, in absence of any external 
currents, given by 
\begin{equation}
J_{x}^{s} =dJ_{c}sin\left(\phi_{2} -\phi _{1} +2\pi \frac{\Phi}{\Phi_{0}} \right) \,\,\, ,
\end{equation}
where $\phi = \phi _{2} -\phi_{1} +2\pi \Phi/\Phi _{0}$ is the phase difference across the junction, 
$\Phi$ the magnetic flux through the junction and $\Phi _{0}$ the flux quantum. The magnetic flux for a 
gaussian magnetization distribution is estimated to be
\begin{equation}
\Phi \approx \mu_{0}\int_{y_{1}}^{y_{2}} \int_{-W}^{W} H_{z}dydz 
=\frac{\mu_{0}\Delta y \lambda_{e} M_{0} }{\sqrt{\pi \alpha}} \int_{0}^{\infty} k_{x} \frac{\exp(-k_{x}^{2}/4\alpha)sin(k_{x}W)}{1+2\lambda _{e}k_{x}}
dk_{x} \,\,\, ,
\end{equation}
where $\Delta y=y_{2}-y_{1}$ is the length of the weak link. 
It is seen that the flux, and therefore also the phase, is dependent on the magnetization, the wall width and the 
penetration depth. 

Let us now consider a normal weak link (i.e. not magnetic), in which the phase can be found by using the 
approach of Ref. \cite{Kogan}. If we position a very thin magnetic structure (as discussed above) directly over the 
predefined weak link, a flux will flow through this junction. Here we may set $\lambda _{e} \rightarrow \infty$, since the 
junction is entirely nonsuperconducting, and this results in
\begin{equation}
\Phi \approx \frac{\mu _{0}\Delta y M_{0}}{2\sqrt{\pi \alpha} }  \int_{0}^{\infty} \exp(-k_{x}^{2}/4\alpha)sin(k_{x}W)
dk_{x} =C\mu _{0}M_{0}\Delta y \,\,\, ,
\end{equation}
where C is an unimportant constant ($C=[\exp(-2)/\sqrt{\pi}]\int _{0}^{2\sqrt{2}} \exp(k_{x}^{2})dk_{x}$), 
and we have assumed that $\alpha =2W^{-2}$. This approach also gives a reasonable description of the case where 
the strength of the magnetic field from the magnetic structure breakes down superconductivity. However, the 
simple treatment given here only accounts for the direct magnetic flux through the junction, and neglects the 
vortices distributed around the weak link. In general, the full 
expression for $\phi$ can not be found explicitly, but can be obtained by 
first determining the vortex interspacing distance by evaluating the expression for the energy 
(i.e. find the zero energy). From the given distribution of vortices one may obtain the magnetic field $H_{vy}$, 
and finally take advantage of the expression $J_{x}^{s} =dJ_{c}sin\phi = -2H_{vy}$ to 
obtain an expression for the phase $\phi$. Such a numerical analysis is outside the scope of this Brief 
Communication. However, let us for completeness discuss some properties of the energy of the system, which is in 
general found by evaluating
\begin{equation}
E = \int _{V}(\frac{1}{2} \mu _{0} \mbox{\boldmath $H$}^{2} +\frac{1}{2} \mu _{0} \lambda ^{2}\mbox{\boldmath $J$}_{s}^{2} -\mu _{0}\mbox{\boldmath $M\cdot H$} )dV \,\,\, .
\end{equation}
Here $\mu _{0}$ is the permeability, and the current due to supercurrents and magnetization gradients can be written as
\begin{equation}
\mbox{\boldmath $J$}=\mbox{\boldmath $J$}_{s} + \mbox{\boldmath $J$}_{m} =\mbox{\boldmath $J$}_{s} + \mbox{\boldmath $\nabla \times M$}  
\end{equation}
Note that the integration over the energy density is taken over the whole space, although the current can only flow in 
the volume of the thin film superconductor. I.e., we do not adopt the usual approach of dividing the 
space into superconducting and nonsuperconducting regions. Thus, we believe that the approach shown below can be applied to a more 
general class of systems, as long as there is no exchange of electrons and spin between the magnetic and superconducting 
structures. In order to obtain a more useful expression 
for the energy, we first transform the part associated with kinetic energy of the superconducting electrons 
\begin{equation}
E_{kinetic} = \frac{1}{2} \mu _{0}\int _{V}\lambda ^{2}\mbox{\boldmath $J$}_{s}^{2}dV = 
\frac{1}{2} \mu _{0}\int _{V}\lambda ^{2}\mbox{\boldmath $J$}_{s} \mbox{\boldmath $\cdot$} \left[\mbox{\boldmath $\nabla$}\times  \left( \mbox{\boldmath $H$} -\mbox{\boldmath $M$} \right) \right]dV \,\,\, .
\end{equation}
Here we have used that $\mbox{\boldmath $\mbox{\boldmath $J$}_{s} =\nabla$}\times (\mbox{\boldmath $H$}-\mbox{\boldmath $M$}) $.
To further transform this integral, we note that a surface integral over the kernel
$\mbox{\boldmath $J$}_{s} \times \left( \mbox{\boldmath $H$} -\mbox{\boldmath $M$}\right)$  
vanishes when the surface is located far from the system. This means that we can write
\begin{equation}
E_{kinetic} = \frac{1}{2} \mu _{0}\int _{V}\lambda ^{2}\mbox{\boldmath $J$}_{s}^{2}dV = 
\frac{1}{2} \mu _{0}\int _{V}\lambda ^{2}  \left( \mbox{\boldmath $H$} -\mbox{\boldmath $M$}\right) \mbox{\boldmath $\cdot$} \mbox{\boldmath $\nabla \times J$}_{s}dV \,\,\, .
\end{equation}
But now it should be remembered that the London equation gives
\begin{equation}
\mbox{\boldmath $\nabla \times J$}_{s} = -\frac{1}{\lambda ^{2}} \mbox{\boldmath $H$} + 
\frac{1}{\lambda ^{2} } \mbox{\boldmath $V$}  \,\,\, ,
\end{equation}
where $\mbox{\boldmath $V$}$ is the vortex source function, which represents all the vortices in the system (In the 
case of Pearl vortices it is simply a sum of delta functions). We obtain
\begin{equation}
E_{kinetic} = \frac{1}{2} \mu _{0}\int _{V}\lambda ^{2}\mbox{\boldmath $J$}_{s}^{2}dV = 
\frac{1}{2} \mu _{0}\int _{V} \left[-\mbox{\boldmath $H$}^{2}+ \mbox{\boldmath $M\cdot H$} +  ( \mbox{\boldmath $H$} -\mbox{\boldmath $M$}) \mbox{\boldmath $\cdot$} \mbox{\boldmath $V$} \right] dV \,\,\, .
\end{equation}
In total, the energy becomes
\begin{equation}
E = \frac{1}{2} \mu _{0}\int _{V} \left[- \mbox{\boldmath $M\cdot H$} +  ( \mbox{\boldmath $H$} -\mbox{\boldmath $M$}) \mbox{\boldmath $\cdot$} \mbox{\boldmath $V$} \right] dV \,\,\, .
\label{general}
\end{equation}
Equation \ref{general} can be applied to systems which are sufficiently local that the surface
corrections can be neglected.
The usefulness of Eq. \ref{general} relies on the fact that for a thin film superconductor it reduces to a two dimensional integral, 
since the vortex source function and the magnetization distribution both are assumed to be located at z=0. Then the Fourier 
analysis of Refs. \cite{Erdin,Helseth} can be applied to obtain simple integrals for the energy.
In their interesting paper Erdin and coworkers demonstrated a different method for calculating the energy within the London approximation\cite{Erdin}.
The technique presented here gives an alternative and perhaps more intuitive route to evaluate the energy. It is 
instructive to divide the energy terms in three different parts. The self-energy of the magnetic structure is
\begin{equation}
E_{m} = -\frac{1}{2} \mu _{0}\int _{V} \mbox{\boldmath $M\cdot H$}_{m} dV\,\,\, ,
\end{equation}
the energy associated with the vortex is
\begin{equation}
E_{v} = \frac{1}{2} \mu _{0}\int _{V} \mbox{\boldmath $H$}_{v} \mbox{\boldmath $\cdot$} \mbox{\boldmath $V$}dV \,\,\, ,
\end{equation}
and the interaction energy 
\begin{equation}
E_{vm} = -\frac{1}{2} \mu _{0}\int _{V} \left[\mbox{\boldmath $M\cdot H$}_{v} -\mbox{\boldmath $H$}_{m}\mbox{\boldmath $\cdot$} \mbox{\boldmath $V$} + \mbox{\boldmath $M$}\mbox{\boldmath $\cdot$} \mbox{\boldmath $V$} \right] dV \,\,\, .
\end{equation}
It is seen that the $E_{vm}$ consists of the interaction between the magnetization and the vortex field, the magnetically generated 
field and the vortex source function and also the interaction between the magnetization and the vortex source function. Note that the 
two first contributions have opposite sign, and their sum is equal to
$-\mu _{0}\int _{V} \mbox{\boldmath $M\cdot H$}_{v} dV$ only in the case when they are equal.
Although this is a reasonable approximation in some particular cases, the above analysis shows 
that it is not correct in general\cite{Helseth,note2,note3}. Moreover, one should take into account the finite distribution of the vortex source 
function.

In conclusion, we have discussed some properties of thin film superconductors in the close vicinity of 
one dimensional magnetic structures. It was found that the magnetic field depends strongly on the magnetic distribution 
function, and simple estimates for the flux through a weak link were derived. Finally, we derived an 
expression for the energy associated with a more general class of systems.

\newpage
\newpage

\newpage


\newpage
 
\begin{references}

\bibitem{Kogan} V.G. Kogan, V.V. Dobrovitsky, Y. Mawatari, R.G. Mints and J.R.
Clem, Phys. Rev. B $\bf{63}$, 144501 (2001).

\bibitem{Sonin} E.B. Sonin, Pis'ma Zh. Tekh. Fiz. $\bf{14}$, 1640 (1988) 
[Sov. Tech. Phys. Lett. $\bf{14}$, 714 (1988) ].

\bibitem{Yi} G. Yi, P.R. Aitchison, W.D. Doyle, J.N. Chapman and C.D.W. Wilkinson, 
J. Appl. Phys. $\bf{92}$, 6087 (2002).

\bibitem{Erdin} S. Erdin, A.F. Kayali, I.F. Lyuksytov, and V.L. Pokrovsky,
Phys. Rev. B $\bf{65}$, 014414 (2002).

\bibitem{Helseth} L.E. Helseth, Phys. Rev. B $\bf{66}$, 104508 (2002).

\bibitem{note1} On p. 1, Eq. 1 in Ref. \cite{Helseth} 
the term containing the vortex source function should be multiplied with $\delta (z)$, and also on 
p. 4 (line 7) the magnetization distribution should be multiplied with $\delta (z)$. These errors 
did not influence the final result.  

\bibitem{note2} Even though the expression for the interaction energy is not complete, it should be pointed 
out that Eqs. 30, 38 and 44 in Ref. \cite{Helseth} may serve as a reasonable starting point 
for computing the energy when the magnetization is directed perpendicular to the film plane.

\bibitem{note3} In Figs. 2 and 4 in Ref. \cite{Helseth} an attractive force is 
defined to be pointing towards the magnetic bubble. Thus, an attractive force 
for negative $\rho$ is defined to be positive and points to the right, whereas an 
repulsive force points to the left. In the case of $\rho >0$, an attractive force is 
negative and points to the left, whereas an repulsive force points to the right. In sum, 
the force is allways attractive or repulsive, depending on the polarities of the magnetic 
bubble as well as the vortex. Although this sign convention is not mathematically pleasing
(because negative $\rho$ is allowed), it is nonetheless allowed physically as long as one 
clearly points out the definition. Unfortunately, this was not done in Ref. \cite{Helseth}.
The author thanks F.M. Peeters for a discussion on this matter.



\end{references}
\end{document}